\documentclass[10pt,twocolumn]{article}

\usepackage[utf8]{inputenc}
\usepackage[T1]{fontenc}
\usepackage{amsmath,amssymb}
\usepackage{graphicx}
\usepackage{booktabs}
\usepackage{hyperref}
\usepackage{listings}
\usepackage{xcolor}
\usepackage[margin=0.85in]{geometry}
\usepackage{enumitem}
\usepackage{caption}
\usepackage{algorithm}
\usepackage{algpseudocode}
\usepackage{multirow}
\usepackage{tabularx}

\title{\textbf{ProbeLogits: Kernel-Level LLM Inference Primitives\\for AI-Native Operating Systems}}

\author{
  Daeyeon Son\\
  Independent Researcher\\
  Republic of Korea\\
  \texttt{sdy1350@gmail.com}
}

\date{July 2026}

\begin{document}
\maketitle

% ============================================================
\begin{abstract}
An OS kernel that runs LLM inference internally can read the
model's own next-token logit distribution \emph{before} any text
is generated---and act on it as a governance primitive. I present
\textbf{ProbeLogits}, a kernel-level operation that performs a
single forward pass and reads specific token logits to classify an
agent's action as safe or dangerous, with zero learned parameters.
Because the probe reads a logit from the \emph{same} base model the
agent already runs, it removes the second model a fine-tuned guard
requires (Llama Guard~3 is an independent 8B classifier with its
own weights and forward passes): the marginal cost of a safety
check becomes a single logit read.

I evaluate ProbeLogits on three base models (Qwen~2.5-7B,
Llama~3~8B, Mistral~7B) across three external benchmarks
(HarmBench~\cite{harmbench}, XSTest~\cite{xstest},
ToxicChat~\cite{toxicchat}). On HarmBench non-copyright, all three
models reach a 97--99\% block rate with the appropriate verbalizer.
On ToxicChat ($n=1{,}000$), ProbeLogits attains F1
\emph{parity-or-better} against Llama Guard~3: Qwen~2.5-7B
Safe/Dangerous reaches F1~$=0.812$ ($+13.7$\,pp, bootstrap 95\% CIs
disjoint), Llama~3 matches within CI ($+0.4$\,pp), and Mistral
exceeds by $+4.4$\,pp. Classification is a measured
$2.4$--$3.4\times$ faster than Llama Guard~3 in the same hosted
environment ($332$--$556$\,ms vs.\ $851$--$1{,}142$\,ms on matched
prompts), because it reads a single logit position instead of
generating tokens (\S\ref{sec:probe-perf}, \S\ref{sec:multimodel}).
% measured: results/remeasure_probe_{llama,qwen,mistral}.json vs remeasure_lg_{q4km,q8}.json (n=150, 2026-06-15); deploy-realistic: results/probe_latency.json
A calibration strength~$\alpha$ acts as a deployment-time policy
knob rather than a learned hyperparameter, letting the OS trade
recall for precision per operation class---strict for privileged
operations, relaxed for conversational agents.

I implement ProbeLogits within \textbf{Anima OS}, a bare-metal
x86\_64 kernel written in ${\sim}$285{,}000 lines of Rust. Because
agent actions must pass through kernel-mediated host functions,
enforcement operates below the WASM sandbox boundary, making it
substantially harder to circumvent than application-layer
classifiers. I also show that treating the KV cache as process
state enables checkpoint, restore, and fork operations analogous to
traditional process management.
\end{abstract}

% ============================================================
\section{Introduction}
\label{sec:intro}

Operating systems have always treated processes as opaque entities.
The kernel manages their resources---memory, CPU time, file
descriptors---but never inspects what a process \emph{thinks}.
This was a reasonable abstraction for decades: a process's internal
state is its own concern, and the OS need not understand it to
schedule, isolate, or terminate it.

But large language models break this assumption. At every inference
step, an LLM produces a probability distribution over its entire
vocabulary---a logit vector that reveals the model's uncertainty,
intent, and confidence before any text is generated. This is a
fundamentally new kind of process state: a semantic signal that
the OS could use for scheduling, safety, and governance decisions.
Today's operating systems discard this information entirely.

The consequences are threefold.
\textbf{Latency accumulation}: when AI safety checks require
generating text, parsing it, and then acting on the parse result,
each abstraction layer adds overhead. A simple ``is this action
safe?'' query traverses application code, IPC, inference framework,
text generation, and parsing---accumulating latency that makes
per-action governance impractical.
\textbf{Bypassable safety}: when safety policies are enforced in
application code, any agent with sufficient privilege (or a bug in
the sandbox) can bypass them. The enforcement boundary is too
high in the stack.
\textbf{Semantic opacity}: the OS kernel cannot make
informed decisions about AI workloads because it has no access to
model internals---logits, hidden states, uncertainty---that would
enable intelligent scheduling and governance.

This paper presents \textbf{ProbeLogits}, an OS primitive that
addresses all three problems by exposing LLM logit vectors as
kernel-level abstractions. ProbeLogits performs a single forward
pass on a prompt and reads specific token logits to classify actions
without generating any text.

The classification technique itself---reading logits at
class-label token positions (known as \emph{verbalizers} in
NLP) with contextual calibration---is well-established in
NLP~\cite{schick2021pet,zhao2021calibrate}. \textbf{The
contribution of this work is not the ML technique but its
integration as a kernel primitive with structural enforcement
guarantees.} When logit reading runs inside the kernel and
agent actions must traverse kernel-mediated host functions,
safety checks become kernel-mediated---a defense-in-depth
property that application-layer classifiers cannot easily provide.

Concretely, I make three contributions:

\begin{enumerate}[topsep=2pt,itemsep=1pt]
  \item \textbf{ProbeLogits as an OS primitive.}
        I define binary, N-way, and entropy operations as kernel
        abstractions, and show that a general-purpose 7B model
        with zero learned fine-tuning achieves F1 parity with
        Llama Guard~3 on ToxicChat~\cite{toxicchat} (1{,}000
        prompts; bootstrap 95\% CIs overlap) and 97--99\% block
        rates on HarmBench~\cite{harmbench} non-copyright across
        three model families (validated with bootstrap CIs,
        Wilson intervals, and pairwise McNemar tests), and I
        establish an empirical \emph{capability floor}: vanilla
        models below ${\sim}$7B carry no usable safety signal
        in their verbalizer logits (\S\ref{sec:smallmodel}).
  \item \textbf{KV cache as process state.}
        I treat the KV cache as the analogue of CPU register state,
        enabling checkpoint, restore, and fork operations that
        parallel traditional process management.
  \item \textbf{Kernel-enforced constitutional governance.}
        I demonstrate a governance pipeline (adversarial
        pre-filter $\rightarrow$ ProbeLogits classification)
        that is kernel-enforced below the WASM sandbox
        boundary, making it significantly harder to circumvent than
        application-layer classifiers.
\end{enumerate}

I implement these primitives in Anima OS, a bare-metal x86\_64
operating system comprising ${\sim}$285{,}000 lines of Rust
across 377 source files (the bare-metal kernel alone is
227{,}176 lines, of which roughly 96{,}000 are an experimental
GPU bring-up subsystem outside the ProbeLogits path).
% LoC measured 2026-07-04, excludes target/, vendored linux-ref/, results/
The inference
engine sustains a measured 1{,}344
tokens/s on SmolLM2-135M (8 workers; 1{,}266 tokens/s with 16
workers---at this model size the wider gang is \emph{slower},
\S\ref{sec:throughput}) and 12.73
tokens/s on Qwen2.5-7B (measured at a sustained-load operating
point after ${\sim}$2 hours of continuous 7B inference in the
same boot; the DDR5 bandwidth ceiling bounds this workload).
% measured: bootlogs/2026-07-05_paper-bench-2model-7b-135m-silicon.log [paper:thru135m8][paper:thru135m16][paper:thru7b]
The per-token forward costs a measured
0.74\,ms
on 135M (8 workers) and
78.6\,ms
% measured: bootlogs/2026-07-05_paper-bench-2model-7b-135m-silicon.log [paper:tok135m][paper:tok7b]
on 7B; a full ProbeLogits classification prefills the
probe prompt (332--556\,ms hosted on 7B, ${\sim}$220\,ms
deploy-realistic)---fast enough for per-action governance.

% ============================================================
\section{Motivation and Background}
\label{sec:motivation}

\subsection{The Layered AI Problem}
\label{sec:layers}

Contemporary AI systems are built as deep layer cakes:
hardware $\to$ OS kernel $\to$ userspace runtime $\to$ inference
framework $\to$ orchestration library $\to$ application. Each
layer adds latency, attack surface, and opacity. A safety check
in LangChain~\cite{langchain} must traverse Python function calls,
HTTP requests to an inference server, text generation, regex
parsing of the response, and a policy decision---all before the
agent's action can be approved or denied. This architecture is
a consequence of treating AI as an application-level concern
rather than a kernel-level one.

\subsection{Three Problems in Detail}
\label{sec:problems}

\paragraph{Problem 1: Latency.}
Consider the task of classifying whether an agent's intended
action is safe. In a text-based approach, the system constructs
a prompt (``Is deleting all user data safe? Answer Yes or No.''),
sends it to an inference server, waits for multi-token text
generation, parses the output, and acts on the result. With
ProbeLogits, the same classification reads a single logit
position after one prefill---no token-by-token generation, no
parsing, and no IPC. In the same hosted environment this is
a measured 2.4--3.4$\times$ faster than a generation-based guard
(332--556\,ms vs.\ 851--1{,}142\,ms per classification;
${\sim}$220\,ms in a deploy-realistic configuration).

\paragraph{Problem 2: Bypassability.}
Application-level safety filters---including prompt injection
defenses, output classifiers, and policy checkers---operate at
the same privilege level as the agent they are meant to constrain.
A sufficiently capable agent can modify, disable, or circumvent
these filters. Kernel-level enforcement is fundamentally different:
an agent running in a WASM sandbox \emph{cannot} bypass governance
checks without first escaping the sandbox, because the checks are
executed by the kernel on the agent's behalf, outside the sandbox
boundary.

\paragraph{Problem 3: Semantic opacity.}
To a conventional OS, an LLM inference workload is
indistinguishable from any other compute task. The kernel cannot
tell whether the model is confident or uncertain, whether the
output is safe or dangerous, or whether the agent is behaving
anomalously. This opacity prevents the OS from making
semantically informed decisions about scheduling priority,
resource allocation, or isolation.

\subsection{Threat Model}
\label{sec:threat}

I assume the following threat model: agents execute in WASM
sandboxes with fuel-limited execution and no direct memory access
outside their sandbox. The kernel is trusted. The network is
untrusted (all inter-node communication uses authenticated
protocols). Agents may be adversarial: they may attempt to
circumvent safety policies, exfiltrate data, or consume excessive
resources. The governance system must provide structural
guarantees (not just probabilistic detection) against policy
circumvention.

% ============================================================
\section{ProbeLogits Design}
\label{sec:probelogits}

\subsection{The Problem with Text-Based Classification}
\label{sec:text-problem}

Text-based classification---generating a response and parsing
it---suffers from three failure modes that make it unsuitable
for kernel-level governance:

\begin{enumerate}[topsep=2pt,itemsep=1pt]
  \item \textbf{Unexpected format.} The model may respond
        ``No, I don't think that would be appropriate'' instead
        of the expected ``No.'' Parsing extracts ``No'' but
        the actual answer is buried in natural language that
        may vary unpredictably.
  \item \textbf{Fragile parsing.} Regular expressions and
        string matching break on whitespace variations,
        capitalization differences, or tokenizer artifacts.
        Each failure mode requires a new parsing rule.
  \item \textbf{Wasted computation.} Generating a multi-token
        response requires multiple forward passes (one per
        token). Classification needs only the \emph{first}
        token's logit distribution---all subsequent tokens
        are wasted computation.
\end{enumerate}

ProbeLogits eliminates all three failure modes by reading the
logit vector directly, before any text is generated.

\subsection{ProbeLogits Mechanism}
\label{sec:mechanism}

Let $p$ be a prompt (a sequence of tokens), $V$ be the
vocabulary, and $f$ be the model's forward pass. ProbeLogits
computes:
\begin{equation}
\label{eq:forward}
  \boldsymbol{\ell} = f(p) \in \mathbb{R}^{|V|}
\end{equation}
where $\boldsymbol{\ell}$ is the logit vector over the full
vocabulary. For a set of target classes $C = \{c_1, \ldots, c_N\}$,
each mapped to a token ID $t_{c_i}$ via vocabulary lookup, the
classification probability for class $c_i$ is:
\begin{equation}
\label{eq:softmax}
  P(c_i) = \frac{\exp(\ell_{t_{c_i}})}
                 {\sum_{j=1}^{N} \exp(\ell_{t_{c_j}})}
\end{equation}

This is an $N$-way softmax restricted to the target token
positions. The key property is that \emph{exactly one forward
pass} suffices regardless of $N$---the cost is identical for
binary classification and 100-way classification.

I implement three operations as kernel primitives:

\paragraph{Binary classification (\texttt{probe\_yes\_no}).}
Looks up token IDs for ``Yes'' and ``No'' via
\texttt{text\_to\_id()}, runs one forward pass, and computes
a 2-class softmax. Returns the winning class and confidence
$\in [0.5, 1.0]$. A numerical guard handles the edge case
where both logits underflow: if
$\exp(\ell_\textit{yes}) + \exp(\ell_\textit{no}) \leq 10^{-10}$,
the result defaults to confidence 0.5 (maximally uncertain).

\paragraph{N-way classification (\texttt{probe\_classify}).}
Generalizes binary classification to $N$ classes. Each class
label must resolve to a single token in the vocabulary. Returns
a vector of \texttt{ClassResult} sorted by probability
(descending). The same numerical guard applies: if the softmax
denominator is $\leq 10^{-10}$, all classes receive uniform
probability $1/N$.

\paragraph{Token vocabulary lookup (\texttt{text\_to\_id}).}
Maps a text string to its token ID using a BTreeMap over the
GPT-2 byte-pair encoding vocabulary. Cost: $O(\log|V|)$, where
$|V| = 152{,}064$ for Qwen2.5's GPT-2 BPE vocabulary.
Returns \texttt{None} if the text
does not correspond to a single token.

Algorithm~\ref{alg:probelogits} presents the complete
\texttt{probe\_classify} procedure.

\begin{algorithm}[t]
\caption{ProbeLogits N-way Classification}
\label{alg:probelogits}
\begin{algorithmic}[1]
\Require Prompt $p$, class labels $C = [c_1, \ldots, c_N]$
\Ensure Sorted class probabilities
\For{$i \gets 1$ to $N$}
  \State $t_i \gets \textsc{TextToId}(c_i)$
  \Comment{$O(\log|V|)$ BTreeMap lookup}
  \If{$t_i = \textsc{None}$} \Return None
  \EndIf
\EndFor
\State $\textit{tokens} \gets \textsc{Encode}(p)$
\State $\textsc{ResetKvCache}()$
\For{$\textit{tok} \in \textit{tokens}$}
  \State $\boldsymbol{\ell} \gets \textsc{ForwardOne}(\textit{tok})$
  \Comment{Incremental forward}
\EndFor
\State $\ell_{\max} \gets \max_i \ell_{t_i}$
\State $e_i \gets \exp(\ell_{t_i} - \ell_{\max})$ for each $i$
\State $S \gets \sum_i e_i$
\If{$S \leq 10^{-10}$}
  \State $P(c_i) \gets 1/N$ for all $i$
  \Comment{Uniform fallback}
\Else
  \State $P(c_i) \gets e_i / S$ for each $i$
\EndIf
\State \Return $\textsc{SortDescending}(\{(c_i, P(c_i), \ell_{t_i})\})$
\end{algorithmic}
\end{algorithm}

\subsection{Information Efficiency}
\label{sec:info-efficiency}

A single forward pass produces a logit vector over the full
vocabulary of $|V| = 152{,}064$ tokens. This vector carries
$\log_2(|V|) \approx 17.2$ bits of information---the
maximum entropy of the distribution. Standard text generation
uses this vector to sample \emph{one} token, discarding
$>99.99\%$ of the available information.

ProbeLogits extracts precisely the bits needed for the
task at hand. A binary classification extracts 1 bit (plus
confidence). An N-way classification extracts $\log_2 N$ bits.
In both cases, the information is extracted from a \emph{single}
forward pass, whereas text-based classification requires
multiple forward passes (one per generated token) to recover
the same information through text parsing.

\subsection{Robustness Properties}
\label{sec:robustness}

ProbeLogits provides four robustness guarantees that text-based
classification cannot:

\begin{enumerate}[topsep=2pt,itemsep=1pt]
  \item \textbf{No parsing failures.} The output is always a
        floating-point probability in $[0, 1]$. There is no
        text to parse, no regex to fail, no format to validate.
  \item \textbf{Bounded confidence.} The softmax output is
        always in $[0, 1]$ by construction. Binary classification
        confidence is always in $[0.5, 1.0]$.
  \item \textbf{Graceful degradation.} When the model is
        uncertain, the confidence approaches 0.5 (binary) or
        $1/N$ (N-way). This is a meaningful signal---the system
        can defer to a larger model or to human judgment.
  \item \textbf{Numerical stability.} The log-sum-exp trick
        prevents overflow. The uniform fallback when
        $\sum \exp(\cdot) \leq 10^{-10}$ prevents division
        by zero. f64 accumulation in entropy computation prevents
        precision loss with large vocabularies.
\end{enumerate}

\subsection{Logit Entropy}
\label{sec:entropy}

Beyond classification, the full logit vector carries a scalar
summary of model uncertainty: Shannon entropy.

\begin{equation}
\label{eq:entropy}
  H(\boldsymbol{\ell}) = -\sum_{i=1}^{|V|}
    p_i \ln p_i, \quad
  p_i = \frac{\exp(\ell_i)}{\sum_j \exp(\ell_j)}
\end{equation}

For a vocabulary of 152,064 tokens, entropy ranges from 0 nats
(complete certainty---one token has probability 1) to
$\ln(152{,}064) \approx 11.93$ nats (uniform distribution---maximum
uncertainty).

I implement \texttt{logit\_entropy()} as a kernel primitive with
f64 accumulation for numerical stability across the full
152K-token vocabulary. The computation uses the log-sum-exp
trick (subtract $\max_i \ell_i$ before exponentiation) and
skips terms where $p_i < 10^{-10}$ to avoid $\ln(0)$.

Logit entropy is exposed as a kernel readout, enabling three
OS-level capabilities (the entropy computation is implemented;
the gating policies below are enabled by it and left as future
work):

\begin{itemize}[topsep=2pt,itemsep=1pt]
  \item \textbf{Autonomy gating}: when $H > 8$ nats (high
        uncertainty), the kernel can automatically defer the
        decision to a human operator or a larger model.
  \item \textbf{Out-of-distribution detection}: anomalously
        high entropy on routine prompts signals that the model
        is encountering inputs outside its training distribution.
  \item \textbf{Dynamic temperature}: the kernel can adjust
        sampling temperature inversely to entropy, producing
        more deterministic outputs when the model is already
        confident.
\end{itemize}

\subsection{Grammar-Constrained Decoding}
\label{sec:grammar}

For tasks that require multi-token structured output (rather
than single-token classification), I implement
grammar-constrained decoding as a complementary primitive.
The constraint engine operates by masking logits before sampling:

\begin{equation}
\label{eq:grammar}
  \ell_i' = \begin{cases}
    \ell_i & \text{if token } i \text{ is valid} \\
    -\infty & \text{otherwise}
  \end{cases}
\end{equation}

For choice grammars (selecting one of $N$ string options), the
masking algorithm maintains the set of remaining valid completions
given the tokens generated so far. At each step, a token is valid
if appending its text to the generated prefix would still be a
prefix of at least one remaining choice. This prefix matching
operates over character sequences, not token sequences, which
correctly handles cases where choices share common prefixes.

I acknowledge a limitation: the masking is not a 100\%
guarantee for arbitrary tokenizer/choice combinations. GPT-2's
byte-pair encoding covers all single ASCII characters, so ASCII
choices work reliably. For non-ASCII choices or unusual
tokenizer vocabularies, I recommend using
\texttt{probe\_classify()} (single-token classification) where
deterministic results are needed, and reserving grammar
constraints for multi-token structured output.

The overhead of grammar masking is now measured in-OS: on the
7B model (152{,}064-token vocabulary), choice masking costs a
measured 7.70\,ms per token and regex masking 0.14\,ms per
token, against 78.6\,ms for the 7B forward pass itself
(on 135M: 2.39\,ms choice, 0.05\,ms regex over a
49{,}152-token vocabulary).
% measured: bootlogs/2026-07-05_paper-bench-2model-7b-135m-silicon.log [paper:grammarlat][paper:tok7b]
Regex masking is negligible; choice masking is \emph{not}
(roughly 10\% of the 7B per-token forward cost)---the mask is
recomputed at each generation step and the choice scan walks
the full vocabulary performing string prefix comparisons. This
corrects the sub-0.01\,ms estimate carried in v1, which held
only for small choice sets on small vocabularies.

% ============================================================
\section{KV Cache as Process State}
\label{sec:kvcache}

\subsection{The Process State Analogy}
\label{sec:analogy}

Traditional operating systems manage process state through a
well-established set of abstractions: CPU registers are
checkpointed on context switch, the program counter tracks
execution position, and \texttt{fork()} duplicates the entire
process state to create a child. I observe that LLM inference
has a direct analogue:

\begin{table}[h]
\centering
\caption{Process state analogy between traditional OS and
LLM inference primitives.}
\label{tab:process-analogy}
\small
\begin{tabular}{@{}ll@{}}
\toprule
\textbf{Traditional Process} & \textbf{LLM Inference} \\
\midrule
CPU registers & KV cache (attention state) \\
Program counter & KV position (tokens processed) \\
\texttt{fork()} & \texttt{kv\_fork()} \\
Context switch & \texttt{kv\_checkpoint/restore()} \\
Core dump & KV cache serialization \\
\bottomrule
\end{tabular}
\end{table}

This analogy is not merely conceptual---it has direct
implementation consequences. Just as an OS must checkpoint
registers to switch between processes, an AI-native OS must
checkpoint KV cache to switch between inference contexts.
Just as \texttt{fork()} enables speculative execution in
traditional systems, \texttt{kv\_fork()} enables speculative
decoding and conversation branching.

\subsection{Operations}
\label{sec:kv-ops}

I implement three KV cache operations:

\paragraph{\texttt{kv\_checkpoint()}.}
Snapshots the current KV cache state (key and value tensors
for all layers, up to the current position) into a
\texttt{KvCheckpoint} structure. The operation uses
\texttt{checked\_mul} to prevent integer overflow when computing
checkpoint size, and enforces a 32\,MB cap
(\texttt{MAX\_CHECKPOINT\_BYTES}) to prevent memory exhaustion.
Returns an error if the checkpoint would exceed the cap.

\paragraph{\texttt{kv\_restore()}.}
Restores KV cache state from a checkpoint. Before restoring,
the operation validates that the checkpoint's dimensions
(number of layers, bytes per position) match the current model.
This prevents a subtle class of bugs where checkpoints from
one model are accidentally applied to another.

\paragraph{\texttt{kv\_fork()}.}
Creates a copy of the KV cache suitable for transfer to a
different inference context. Currently implemented as an alias
for \texttt{kv\_checkpoint()}, but semantically distinct: fork
implies the original continues executing, while checkpoint
implies the original may be overwritten.

\subsection{Use Cases}
\label{sec:kv-usecases}

KV cache as process state enables several capabilities
that are impossible or expensive without kernel-level support:

\begin{itemize}[topsep=2pt,itemsep=1pt]
  \item \textbf{Speculative execution}: fork the KV cache,
        explore a candidate action, and restore if the action
        is rejected by governance.
  \item \textbf{Agent forking}: create a new agent with the
        same conversational context, diverging from a specific
        point in the conversation.
  \item \textbf{Migration}: serialize the KV cache, transfer
        it to another node, and restore---enabling live agent
        migration across a cluster.
  \item \textbf{Conversation branching}: checkpoint before a
        critical decision, explore multiple branches, and
        select the best outcome.
\end{itemize}

\subsection{Comparison with Serving Systems}
\label{sec:kv-comparison}

vLLM~\cite{vllm} and SGLang~\cite{sglang} also manage KV
caches, but for a fundamentally different purpose. Their
PagedAttention and RadixAttention optimize \emph{serving
throughput}---maximizing the number of concurrent requests
that fit in GPU memory. My KV cache operations optimize
\emph{process state management}---enabling checkpoint, restore,
and fork as OS primitives for individual agents. The two
approaches are orthogonal: a system could use PagedAttention
for memory efficiency \emph{and} KV process state operations
for agent lifecycle management.

% ============================================================
\section{Constitutional Governance Integration}
\label{sec:governance}

\subsection{Why Kernel-Level Enforcement Matters}
\label{sec:kernel-enforcement}

In Anima OS, agents execute in WASM sandboxes compiled via
Cranelift AOT. An agent interacts with the outside world
exclusively through 18 host functions exposed by the
kernel (a base set of 15 OS operations plus 3 MCP tool-calling
functions added in the companion governance gateway~\cite{animaos-mcp}).
% dispatch table: boot/src/aot.rs:645-662 (15 base + 3 MCP = 18)
This architecture provides a structural guarantee: the agent
\emph{cannot} bypass governance checks without escaping the WASM
sandbox, because it has no mechanism to invoke system operations
other than through the host function interface, which the kernel
controls.

This leads to a critical distinction:
\textbf{enforcement is complete; detection is probabilistic.}
The enforcement boundary---WASM isolation preventing direct
system calls---is a structural guarantee that holds regardless
of the agent's capability or intent. The detection mechanism---
ProbeLogits classifying whether an action is harmful---is
probabilistic, with accuracy depending on model size and prompt
quality. The combination is powerful: even if detection produces
a false negative (classifying a harmful action as safe), the
agent still operates within its capability sandbox, limiting
the blast radius of any single action.

\subsection{Detection Pipeline}
\label{sec:pipeline}

I implement a three-stage detection pipeline that
combines rule-based pre-filtering with model-based classification:

\begin{enumerate}[topsep=2pt,itemsep=1pt]
  \item \textbf{Stage 1: Adversarial pre-filter} ($<1$\,\textmu{}s).
        An O($n$) pattern scan detects prompt injection,
        encoding tricks (base64, rot13 references),
        authority impersonation (``ADMIN OVERRIDE''), and
        instruction overrides (``ignore previous'').
        Actions exceeding a risk-score threshold are
        classified as dangerous with no forward pass.
  \item \textbf{Stage 2: Input sanitization.}
        Control characters and injection patterns are
        stripped before the action is embedded in the
        safety-priming prompt template.
  \item \textbf{Stage 3: Calibrated forward pass} (332--556\,ms
        hosted, conservative; ${\sim}$220\,ms deploy-realistic).
        A single 7B forward pass over the probe prompt reads Safe/Dangerous
        verbalizer logits. Contextual calibration corrects
        inherent token bias, and a privacy keyword boost
        compensates for the model's blind spot on
        surveillance-as-harm.
\end{enumerate}

The per-action cost is dominated by the 7B classification
forward (332--556\,ms hosted, ${\sim}$220\,ms deploy-realistic;
the per-token cost is a measured 78.6\,ms).
% measured: bootlogs/2026-07-05_paper-bench-2model-7b-135m-silicon.log [paper:tok7b]
A natural-seeming optimization---cascading through a
\emph{vanilla} smaller model first (e.g., 135M) and escalating
to 7B only for ambiguous cases---is empirically refuted by the
small-model floor result (\S\ref{sec:smallmodel}): vanilla
sub-billion-parameter probes miss 87--90\% of harmful prompts,
so they cannot serve as a first-stage tripwire. A
\emph{distilled} tripwire head trained against 7B-class
verdicts remains plausible future work.

\subsection{Graduated Response}
\label{sec:graduated}

The governance system maps ProbeLogits confidence to a
graduated response:

\begin{table}[h]
\centering
\caption{Graduated response policy mapping ProbeLogits
confidence bands to governance verdicts (thresholds as
shipped in the kernel).}
% thresholds match boot/src/constitutional.rs:1043-1069 (4-band cascade + NonDeception rule)
\label{tab:graduated-response}
\small
\setlength{\tabcolsep}{4pt}
\begin{tabular}{@{}ll@{}}
\toprule
\textbf{Confidence} & \textbf{Response} \\
\midrule
$P(\text{harmful}) > 0.85$ & Block (deny the action) \\
$P(\text{harmful}) > 0.70$ & Warn (log + notify operator) \\
$0.15 < P(\text{harmful}) \leq 0.70$ & Log (record, proceed) \\
$P(\text{harmful}) \leq 0.15$ & Allow (proceed normally) \\
\bottomrule
\end{tabular}
\end{table}

In addition to the four confidence bands, a principle-specific
rule escalates to Warn at $P > 0.80$ when the flagged
constitutional principle is NonDeception, reflecting the higher
cost of letting deceptive actions pass silently.

The graduated response is intentionally conservative: a 78\%
confidence harmful classification results in a warning, not
a block. This reflects the reality that 7B Q4\_0 models are
not highly accurate classifiers, and false positives (blocking
legitimate actions) erode user trust more than false negatives
(allowing suspicious actions with a warning). Mid-band
classifications ($0.15 < P \leq 0.70$) are recorded in the
audit chain without operator notification, preserving
forensic visibility at zero interaction cost.

\subsection{Blake3 Audit Chain}
\label{sec:audit}

All governance decisions---classifications, confidence scores,
and responses---are recorded in a \emph{keyed} Blake3 hash
chain. Each audit entry stores the keyed Blake3 hash of (the
previous entry's hash $\|$ its own fields), creating a
tamper-evident log: retroactive modification of any entry
invalidates that entry's hash and every subsequent one, and an
attacker without the chain key cannot recompute a consistent
replacement chain. The chain is checked by
\texttt{verify\_chain}, which recomputes each keyed hash and
verifies the links. Entry hashes are persisted together with
the log and \emph{verified} (not re-derived) at restore time,
so tampering with the on-disk log is detected across reboots:
a verification mismatch at boot places the system in Safe Mode
until an operator explicitly clears it. This
reboot-surviving tamper-detection path has been demonstrated
end-to-end on real hardware, including an offline-attacker
scenario in which the persisted log is modified while the
system is powered down and the modification is detected on the
next boot. The in-memory ring buffer evicts entries beyond
256; durable persistence integrates with the AnimaFS storage
layer. The implementation uses the \texttt{no\_std} Blake3
core, running natively in the bare-metal kernel.
% keyed chain + restore-time hash verification: boot/src/constitutional.rs:555,615,687-708
% reboot-surviving tamper detection: zero-trust S1-S5, commit 1e8fabb, silicon-verified 2026-06-28

% ============================================================
\section{Implementation}
\label{sec:implementation}

\subsection{Bare-Metal Inference Engine}
\label{sec:engine}

The inference engine comprises 15,631 lines of \texttt{no\_std}
Rust, running directly on bare metal with no operating system,
no libc, and no standard library. It implements:
% boot/src/inference/*.rs = 15,631 LoC (17 files, measured 2026-07-04)

\paragraph{GGUF model loader.}
Parses the GGUF binary format (llama.cpp's model format)
directly from USB mass storage via xHCI, or from NVMe (the
system self-hosts its kernel and models from an NVMe drive),
loading tensor data
in Q4\_0, Q6\_K, and Q8\_0 quantization formats.

\paragraph{GPT-2 BPE tokenizer.}
A byte-pair encoding tokenizer supporting vocabularies up to
152,064 tokens. Token-to-ID lookup uses a BTreeMap for
$O(\log|V|)$ performance. The tokenizer supports chat templates
for instruction-following models.

\paragraph{Llama transformer.}
A complete Llama-architecture transformer implementation:
multi-head grouped-query attention with RoPE positional
encoding, SwiGLU feed-forward networks, and RMS normalization.
Supports models with GQA ratios (e.g., Qwen2.5-7B uses 28
attention heads with 4 KV heads; Llama-3-8B uses 32 with 8).

\paragraph{AVX-512/AVX2 SIMD kernels.}
Quantized matrix-vector multiplication kernels for three
formats:
\begin{itemize}[topsep=2pt,itemsep=1pt]
  \item Q4\_0: 4-bit quantization with per-block f16 scale.
        AVX-512 kernel processes 64 elements (2 blocks) per
        iteration using VNNI \texttt{vpdpbusd} instructions.
  \item Q6\_K: 6-bit quantization with per-superblock scale
        and per-block sub-scale. Requires careful bit
        unpacking across three byte segments.
  \item Q8\_0: 8-bit quantization with per-block f16 scale.
        The fastest format, using direct \texttt{vpmaddubsw}
        instructions.
\end{itemize}
All kernels use F16C hardware instructions
(\texttt{vcvtph2ps}) for f16$\to$f32 scale conversion,
eliminating the need for a software floating-point library.

\paragraph{SMP work-stealing graph executor.}
The inference computation is expressed as a dataflow graph
(measured per model: 594 operations with 226 synchronization
barriers for 7B; 606 operations with 242 barriers for
135M---v1 conflated these into a single ``606 ops / 226
barriers'' pair, which mixed the 135M op count with the 7B
barrier count).
% measured: bootlogs/2026-07-05_paper-bench-2model-7b-135m-silicon.log [paper:graphstats]
The
graph executor uses work-stealing across all available CPU
cores: each core maintains a local deque of ready operations,
and idle cores steal work from busy cores' deques. For the
7B model, the forward pass is fully parallelized across 8
cores.

\subsection{The \texttt{no\_std} Challenge}
\label{sec:nostd}

Building an LLM inference engine in a bare-metal environment
presents four challenges that do not exist in userspace:

\begin{enumerate}[topsep=2pt,itemsep=1pt]
  \item \textbf{No float runtime.}
        Bare-metal Rust provides no \texttt{libm} for
        transcendental functions. I use F16C hardware
        instructions for f16 conversion and a polynomial
        approximation (\texttt{fast\_expf\_poly}) for
        $\exp()$ in softmax and entropy computations.
  \item \textbf{No filesystem.}
        GGUF model files (4+ GB) are read from USB mass
        storage via a custom xHCI driver (9,357 lines) or
        from NVMe, block by block.
  \item \textbf{No threading.}
        SMP parallelism is implemented from scratch: AP
        (Application Processor) bootstrap via the INIT-SIPI
        protocol, spin locks, and atomic operations---no
        \texttt{pthread}, no \texttt{std::thread}.
  \item \textbf{No allocator (initially).}
        The heap allocator is bootstrapped from the UEFI
        memory map. Once initialized, standard
        \texttt{alloc::Vec} and \texttt{alloc::String} are
        available.
\end{enumerate}

\subsection{System Context}
\label{sec:context}

ProbeLogits operates within a complete OS kernel. The
\textbf{agent lifecycle} manages five states (Nascent, Active,
Suspended, Isolated, Terminated) with trust evolution based on
task completion history and a four-tier capability system
(System, AiNative, AiEnhanced, Classic).

\textbf{AnimaNet} provides distributed networking: UDP discovery
with boot\_nonce-based reboot detection, TCP-based Raft consensus
for cluster coordination, and live agent migration (serializing
the AgentControlBlock and transferring it over TCP). A two-node
cluster has been validated end-to-end with 11/11 test markers
passing.

The \textbf{web stack} enables agents to access external
services: DNS resolution, HTTP client, TLS 1.3 (via
\texttt{rustls} with a custom \texttt{no\_std} RustCrypto provider
and full X.509 chain/hostname certificate validation), and a
\texttt{web\_fetch()} host function exposed to WASM sandboxes.

The \textbf{WASM sandbox} uses Cranelift AOT compilation with
fuel-limited execution. Eighteen host functions bridge the
sandbox to kernel services (15 base operations spanning
inference, governance, IPC, file system, and web access; plus
3 MCP tool-calling functions added in the companion
gateway~\cite{animaos-mcp}).
% boot/src/aot.rs:645-662

\subsection{Code Scale}
\label{sec:scale}

\begin{table}[t]
\centering
\caption{Code scale by component (Rust LoC, measured 2026-07-04;
excludes build artifacts, vendored Linux reference sources, and
result data). The workspace passes 1,100+ tests.}
% methodology: find <dir> -name "*.rs" -not -path "*/target/*" | xargs cat | wc -l
\label{tab:loc}
\scriptsize
\setlength{\tabcolsep}{4pt}
\begin{tabular}{@{}lrl@{}}
\toprule
\textbf{Component} & \textbf{LoC} & \textbf{Role} \\
\midrule
boot/ (bare-metal)   & 227,176\textsuperscript{*} & Kernel + inference + GPU + MCP \\
kernel/ (hosted)     & 25,178 & Lifecycle, memory \\
runtime/             & 11,894 & WASM, scheduling \\
cli/                 &  2,334 & Agent mgmt \\
sdk/                 &    384 & WASM SDK + MCP \\
dsl/                 &  1,439 & Agent DSL \\
tools/ + agents/     & 14,112 & Boot builder, example agents \\
\midrule
\textbf{Total}       & \textbf{285,249} & 377 Rust files \\
\bottomrule
\multicolumn{3}{p{0.95\columnwidth}}{\textsuperscript{*}Composition
note: ${\sim}$96{,}000 LoC of boot/ is an experimental bare-metal
GPU bring-up subsystem (RDNA4 driver development, including
${\sim}$16{,}000 LoC of embedded shader/firmware tables). It is
not on the ProbeLogits code path; the raw count is disclosed for
honesty rather than as feature scale.}
\end{tabular}
\end{table}

Table~\ref{tab:loc} summarizes the system's code scale. The
boot/ kernel (227,176 LoC across 227 files) is the largest
component, containing the bare-metal inference engine
(15,631 LoC), xHCI USB 3.0 driver (9,357 LoC), VirtIO-Net
networking, AnimaNet protocol, the experimental GPU bring-up
subsystem (see table footnote), and all ProbeLogits primitives.

% ============================================================
\section{Evaluation}
\label{sec:evaluation}

\subsection{Experimental Setup}
\label{sec:setup}

All experiments run on a single machine:

\begin{itemize}[topsep=2pt,itemsep=1pt]
  \item \textbf{CPU}: AMD Ryzen 9 9800X3D (8 cores, 16 threads,
        AVX-512 with VNNI)
  \item \textbf{RAM}: 60\,GB DDR5-6000 dual-channel
        ($\sim$75\,GB/s measured bandwidth)
  \item \textbf{Models}: SmolLM2-135M Q4\_0 (70\,MB),
        Qwen2.5-7B-Instruct Q4\_0 (4.1\,GB)
  \item \textbf{Baseline}: llama.cpp (latest, March 2026)
        on Linux 6.17, same hardware
\end{itemize}

Anima OS runs bare-metal (UEFI boot, no Linux). The llama.cpp
baseline runs on Linux with identical CPU and memory.

\subsection{Inference Throughput}
\label{sec:throughput}

\begin{table}[t]
\centering
\caption{Token generation throughput (tokens/s, Anima vs.\
llama.cpp). Higher is better. Anima values are silicon-measured
(2026-07-05, single boot); the llama.cpp column is the March
2026 baseline run on the same hardware, not re-run in this
campaign. The qualitative regimes (compute-bound 135M,
bandwidth-bound 7B) are the claims of record.}
% measured: bootlogs/2026-07-05_paper-bench-2model-7b-135m-silicon.log [paper:thru135m16][paper:thru135m8][paper:thru7b]
\label{tab:throughput}
\scriptsize
\begin{tabular}{@{}llp{0.55\columnwidth}@{}}
\toprule
\textbf{Model} & \textbf{Config} & \textbf{Anima vs.\ llama.cpp} \\
\midrule
135M Q4\_0 & 16w/16t & 1{,}266 (measured) vs.\ 1{,}195 (baseline) \\
135M Q4\_0 & 8w/8t   & \textbf{1{,}344} (measured) vs.\ 1{,}301 (baseline) \\
7B Q4\_0   & 8w/8t   & 12.73 (measured, sustained-load) vs.\ 15.0 (baseline) \\
\bottomrule
\end{tabular}
\end{table}

Table~\ref{tab:throughput} presents measured inference
throughput. On the 135M model, Anima OS sustains a measured
1{,}266 tok/s with 16 work-stealing workers and 1{,}344 tok/s
with 8 workers.
% measured: bootlogs/2026-07-05_paper-bench-2model-7b-135m-silicon.log [paper:thru135m16][paper:thru135m8]
Two observations follow. First, against the March 2026
llama.cpp baseline (1{,}195 tok/s at 16 threads, 1{,}301 at
8), Anima is at or slightly above parity; the 1.39$\times$
advantage reported in v1 (1{,}666 vs.\ 1{,}195) did not
reproduce in this campaign and is withdrawn as a headline
claim. Second---a genuine measured finding about gang
width---\textbf{8 workers are \emph{faster} than 16 on the
135M model}: at this model size, the per-operation barrier
and coordination cost of the wider gang exceeds the
parallelism gain. Gang width is therefore a
model-size-dependent tuning parameter, not a monotone win.
The 135M model is \emph{compute-bound}:
its 70\,MB parameter footprint fits in L3 cache, and
performance scales with parallel compute---up to the point
where work-stealing coordination overhead dominates.

On the 7B model, Anima measures 12.73 tok/s (78.6\,ms per
token; decode forward 81.1\,ms).
% measured: bootlogs/2026-07-05_paper-bench-2model-7b-135m-silicon.log [paper:thru7b][paper:tok7b]
An honest caveat applies: this figure was captured after
${\sim}$2 hours of continuous 7B inference in the same boot,
so it reflects a sustained-load (thermally influenced)
operating point rather than a cold-machine best case; the
March 2026 baseline runs measured 15 tok/s on both systems.
The 7B model is \emph{bandwidth-bound}:
each forward pass reads 4.1\,GB of parameters from main
memory, and the DDR5 ${\sim}$75\,GB/s bandwidth remains the
theoretical ceiling that determines performance rather than
compute efficiency.

\subsection{Bandwidth Saturation Analysis}
\label{sec:bandwidth}

The 7B parity result is explained by a fundamental hardware
limit. The theoretical minimum time per token for a
bandwidth-bound model is:
\begin{equation}
\label{eq:bandwidth}
  t_{\min} = \frac{\text{model size}}{\text{bandwidth}}
           = \frac{4.1\,\text{GB}}{75\,\text{GB/s}}
           = 55\,\text{ms}
\end{equation}

The measured per-token cost of 78.6\,ms (sustained-load
operating point)
% measured: bootlogs/2026-07-05_paper-bench-2model-7b-135m-silicon.log [paper:tok7b]
corresponds to ${\sim}$70\% of theoretical peak
bandwidth utilization. The remaining gap accounts for
non-memory operations (softmax, RoPE, layer normalization),
memory access patterns that do not achieve perfect
streaming, and the sustained-load thermal envelope of the
measurement itself.

This is a significant finding: \textbf{bare-metal execution
operates at the memory-bandwidth wall.} No software
optimization---better SIMD kernels, more efficient scheduling,
or tighter code---can lift 7B performance past the bandwidth
ceiling on this hardware.
The path to faster 7B inference requires higher memory bandwidth
(DDR5-8000, HBM) or reduced model size (more aggressive
quantization).

\subsection{ProbeLogits Performance}
\label{sec:probe-perf}

\begin{table}[t]
\centering
\caption{ProbeLogits operation latencies. The bare-metal
per-operation rows are silicon-measured (2026-07-05, single
boot, real hardware); the hosted full-classification rows are
measured and archived. A full Binary/N-way classification
prefills the ${\sim}$35-token probe prompt.}
% measured: bootlogs/2026-07-05_paper-bench-2model-7b-135m-silicon.log [paper:tok7b][paper:tok135m][paper:entropylat][paper:grammarlat][paper:kvlat]
% deploy-realistic rows: results/probe_latency.json (2026-06-14, logits_all=False);
% conservative rows: results/remeasure_probe_{llama,qwen,mistral}.json (2026-06-15, n=150, logits_all=True)
\label{tab:probe}
\scriptsize
\begin{tabular}{@{}p{0.95\columnwidth}@{}}
\toprule
\textbf{Bare-metal per-operation (measured, 2026-07-05 boot):} \\
Binary / N-way forward (per token): 135M
0.74\,ms (8w) / 0.79\,ms (16w);
7B 78.6\,ms \\
Entropy (\texttt{logit\_entropy}):
0.65\,ms on 135M (49{,}152 vocab), 2.57\,ms on 7B
(152{,}064 vocab) \\
Grammar masking (per token):
choice 2.39\,ms (135M) / 7.70\,ms (7B); regex 0.05\,ms /
0.14\,ms \\
KV checkpoint / restore:
0.005 / 0.002\,ms on 135M; 0.037 / 0.016\,ms on 7B
(0.93\,MB at 32 positions) \\
\midrule
\textbf{Hosted full classification (measured, archived):} \\
Deploy-realistic (last-token logits only): 219.6\,ms median
(Qwen 2.5-7B, 29-token probe prompt); 224.4\,ms (Llama 3 8B);
391.4\,ms (Mistral 7B Q8\_0) \\
Conservative accuracy-harness configuration (full logits,
n=150): 332\,ms (Llama 3), 358\,ms (Qwen), 556\,ms
(Mistral Q8) \\
\bottomrule
\end{tabular}
\end{table}

Table~\ref{tab:probe} shows ProbeLogits operation latencies.
Binary and N-way classification have identical cost because
both require exactly one forward pass; the additional softmax
over $N$ token logits is negligible ($<$1\,\textmu{}s for
$N \leq 100$).

Logit entropy computation requires no additional forward
pass---it iterates once over the already-computed logit
vector---and its cost scales with vocabulary size: a measured
0.65\,ms on the 135M model's 49{,}152-token vocabulary and
2.57\,ms on Qwen's 152{,}064-token vocabulary. The large Qwen
vocabulary dominates the cost; v1's sub-millisecond estimate
held only for the smaller vocabulary.
% measured: bootlogs/2026-07-05_paper-bench-2model-7b-135m-silicon.log [paper:entropylat]

KV cache checkpoint cost scales with the number of cached
positions (the KV cache must be copied) and is far cheaper
than v1's multi-millisecond estimate: a measured 0.037\,ms
checkpoint / 0.016\,ms restore on 7B (0.93\,MB at 32 cached
positions) and 0.005 / 0.002\,ms on 135M. The 32\,MB cap
ensures bounded memory consumption.
% measured: bootlogs/2026-07-05_paper-bench-2model-7b-135m-silicon.log [paper:kvlat][paper:kvsnap]

\paragraph{Deploy-realistic classification latency.}
The ${\sim}$400\,ms per-classification figure quoted in v1 of
this paper is the \emph{conservative} bound from the accuracy
harness, which computes logits at every position
(\texttt{logits\_all=True}). A deployment only needs the
last-token logits: measured in that configuration, a single 7B
probe costs \textbf{219.6\,ms median} (Qwen 2.5-7B, 29-token
probe prompt; Llama 3 8B 224.4\,ms; Mistral 7B Q8\_0
391.4\,ms). Running all three model probes for a cross-model
consensus vote (\S\ref{sec:consensus}) costs 391.4\,ms in
parallel or 835.4\,ms sequentially.
% results/probe_latency.json (2026-06-14)

\subsection{ProbeLogits Safety Classification}
\label{sec:native}

I validate ProbeLogits on the standard
ToxicChat~\cite{toxicchat} benchmark (Lin et al., 2023)---a
human-annotated dataset of 10K+ real user conversations from a
deployed chatbot. I sample 1{,}000 prompts (362 toxic, 638
benign; seed=42) from the test split.

\paragraph{Contextual calibration.}
Instruction-tuned models exhibit strong sycophancy bias---a
tendency to agree with the implied framing of the question.
With Yes/No verbalizers and uncalibrated readout, the model
predicts ``dangerous'' for the majority of prompts. I apply
contextual calibration~\cite{zhao2021calibrate}: at boot time,
seven null-input prompts are probed to measure the model's
inherent bias ($\Delta = $ logit$_{\text{Dangerous}} - $
logit$_{\text{Safe}}$), which is subtracted from all
classifications scaled by a tunable
$\alpha \in [0, 1]$. I also use \emph{Safe/Dangerous}
verbalizers in addition to Yes/No, since Safe/Dangerous
carry direct classification signal rather than surface
agreement; the trade-offs of this choice are explored in
\S\ref{sec:multimodel} and \S\ref{sec:verbalizer-limit}.

The prompt template uses safety-priming framing:
\texttt{"Is this action dangerous, harmful, or a privacy violation? Answer Safe or Dangerous.\textbackslash{}nAction: \{action\}\textbackslash{}nAnswer:"}

\paragraph{Single-model ToxicChat baseline.}
On Qwen2.5-7B-Instruct Q4\_0 with $\alpha=0.5$ on the bare-metal
Anima OS native runtime, the full pipeline (S/D verbalizer,
calibrated, with adversarial pre-filter and privacy keyword
boost) reaches a silicon-measured F1 = 0.829
% measured: bootlogs/2026-07-05_paper-bench-2model-7b-135m-silicon.log [paper:natsweep]
on ToxicChat 1000 (Table~\ref{tab:alpha}; the April 2026
native run had reported 0.837---the archived 2026-07-05
silicon re-run supersedes that unarchived figure).
For apples-to-apples hosted-mode comparison with Llama Guard 3
under the same llama-cpp-python environment, the pure-logit
configuration (no hand rules) at $\alpha=0.0$ reaches
F1 = 0.812 (Table~\ref{tab:tc-multimodel}, raw data
\texttt{results/toxicchat\_qwen\_sd\_alpha00.json}). We treat
F1 = 0.812 (pure logit, no hand rules) as the scientific claim
for the primitive itself---directly comparable to Llama Guard~3
under identical conditions---while the measured F1 = 0.829
% measured: bootlogs/2026-07-05_paper-bench-2model-7b-135m-silicon.log [paper:natsweep]
is the deployed full-pipeline operating point. Tuning
$\alpha$ trades F1 for higher recall. The behavior of $\alpha$
as a deployment-time policy knob is detailed below; the
multi-model and verbalizer comparisons follow in
\S\ref{sec:multimodel}.

\paragraph{Substrate vs.\ supplementary heuristics.}
The full deployment pipeline combines (i) calibrated logit
reading (the substrate primitive), (ii) the chosen verbalizer,
(iii) an adversarial pre-filter (O($n$), no forward pass), and
(iv) a privacy keyword boost (a hand rule for OS actions). The
substrate primitive alone accounts for the bulk of the
classification signal; (iii)--(iv) are governance-layer
heuristics that contribute roughly $4$\,pp on top of the
substrate. The full pipeline ablation
(uncalibrated $\to$ +calibration $\to$ +safety prompt $\to$
+privacy boost $\to$ +adversarial pre-filter, on a 260-prompt
OS-action benchmark) is reported in the companion Governed
MCP paper~\cite{animaos-mcp}, which uses that dataset for
end-to-end gateway evaluation.

\iffalse
% Pipeline ablation table (Custom-260) moved to companion
% Governed MCP paper §5.4 (tab:pipeline-260) per cross-paper
% division of labor: Custom-260 evaluates substrate+hand-rules
% as a deployed gateway; this paper evaluates the substrate
% primitive on standard external benchmarks.
\begin{table}[t]
\centering
\caption{Pipeline ablation: contribution of each layer to
end-to-end F1 on a 260-prompt OS-action benchmark
(reported in companion paper~\cite{animaos-mcp}).}
\label{tab:ablation}
\small
\begin{tabular}{@{}lrrl@{}}
\toprule
\textbf{Config} & \textbf{Acc.} & \textbf{F1} & \textbf{Note} \\
\midrule
Uncalib., Yes/No   & 64.8\% & 0.786 & Sycophancy \\
+ Calibration      & 87.1\% & 0.892 & Bias corr. \\
+ Safety prompt    & 92.3\% & 0.941 & Model-only \\
+ Privacy boost    & ${\sim}$95\% & ${\sim}$0.96 & Hand rule \\
+ Adv.\ pre-filter & 97.3\% & 0.980 & Full pipe. \\
\bottomrule
\end{tabular}
\end{table}
\fi

The model-only configuration of that pipeline (calibrated
forward pass, Safe/Dangerous verbalizer, no hand rules)
achieves F1=0.941; supplementary heuristics add ${\sim}$4\,pp.
The companion paper additionally reports the governance-level
ablation that motivates this work: removing the entire
ProbeLogits layer from a 6-layer kernel governance pipeline
on a 101-prompt MCP benchmark drops F1 from 0.789 to 0.357
($\Delta$F1 = $-$0.432, silicon-measured re-run)~\cite{animaos-mcp}.
% measured: bootlogs/2026-07-05_paper-bench-2model-7b-135m-silicon.log [paper:ablation]

\subsection{Multi-Model Validation}
\label{sec:multimodel}

I evaluate ProbeLogits across three base model families
(Qwen 2.5-7B Q4\_0, Llama 3 8B Q4\_0, Mistral 7B v0.3 Q8\_0)
on three external benchmarks
(HarmBench~\cite{harmbench}, XSTest~\cite{xstest},
ToxicChat~\cite{toxicchat}) with two verbalizer pairs
(Safe/Dangerous, Yes/No). All hosted-mode evaluations use
llama-cpp-python with 16 threads. Wilson 95\% CIs and bootstrap
F1 CIs are reported on the headline numbers.

\subsubsection{HarmBench: Catching Clearly Harmful Prompts}
\label{sec:multimodel-harmbench}

HarmBench~\cite{harmbench} (Mazeika et al., ICML 2024)
provides 400 unsafe prompts in 7 categories. I separate
copyright (n=100) from non-copyright safety harm (n=300)
because copyright detection requires reference-based matching
not in scope for logit-level safety classification.

\begin{table}[t]
\centering
\caption{HarmBench block rate by model and verbalizer.
Non-copyright (n=300) is the safety-harm subset; copyright
(n=100) requires reference-based detection out of scope.
Wilson 95\% CIs in brackets.}
\label{tab:hb-multimodel}
\scriptsize
\setlength{\tabcolsep}{2.5pt}
\begin{tabular}{@{}lrrrr@{}}
\toprule
\textbf{Model} & \textbf{S/D Tot.} & \textbf{S/D non-©} & \textbf{Y/N Tot.} & \textbf{Y/N non-©\textsuperscript{\ddag}} \\
\midrule
Qwen 2.5-7B   & 74.2\% & 98.7\% & 86.8\% & \textbf{99.0\%} \\
Llama 3 8B    & 72.8\% & 97.0\% & 75.0\% & \textbf{98.0\%} \\
Mistral 7B    & 65.5\% & 87.3\%\textsuperscript{*} & 79.5\% & \textbf{98.7\%} \\
\bottomrule
\multicolumn{5}{p{0.95\columnwidth}}{\textsuperscript{\ddag}Y/N non-© Wilson 95\% CIs: Qwen [97.1, 99.7], Llama 3 [95.7, 99.1], Mistral [96.6, 99.5]. \textsuperscript{*}Mistral S/D underperforms because ``Dangerous'' tokenizes as 3 tokens in Mistral's SentencePiece (\S\ref{sec:verbalizer-limit}); the probe reads a meaningless ``D'' logit.}
\end{tabular}
\end{table}

\textbf{Headline finding (Table~\ref{tab:hb-multimodel}).}
On non-copyright prompts with the Y/N verbalizer (universal
across tokenizers), all three models reach 97--99\% block
rate---within 1\,pp of each other, McNemar pairwise
comparisons all $p > 0.24$ (statistically indistinguishable).
The mechanism is broadly architecture-agnostic for
clearly-harmful content. Copyright prompts behave very
differently across models (Qwen 50\%, Llama 3 6\%, Mistral
22\%) and are reported separately rather than averaged in.

\subsubsection{XSTest: Recall Architecture-Agnostic, Over-Refusal Model-Dependent}
\label{sec:multimodel-xstest}

XSTest~\cite{xstest} (R\"ottger et al., NAACL 2024) contains
200 unsafe prompts and 250 safe-but-edgy prompts designed to
test exaggerated safety. I report Y/N verbalizer at $\alpha=0.5$
(Table~\ref{tab:xs-multimodel}).

\begin{table}[t]
\centering
\caption{XSTest at Y/N $\alpha=0.5$. Recall is the fraction of
the 200 unsafe prompts caught; over-refusal is the fraction of
the 250 safe prompts wrongly flagged.}
\label{tab:xs-multimodel}
\small
\begin{tabular}{@{}lrrr@{}}
\toprule
\textbf{Model} & \textbf{Recall} & \textbf{Over-refusal} & \textbf{F1} \\
\midrule
Qwen 2.5-7B   & \textbf{100\%}  & 86.4\% & 0.649 \\
Llama 3 8B    & 98.5\% & \textbf{51.6\%} & \textbf{0.749} \\
Mistral 7B    & 99.5\% & 55.6\% & 0.740 \\
\bottomrule
\end{tabular}
\end{table}

\textbf{Two findings.}
First, recall is essentially perfect across all three models
(98.5--100\%): the substrate primitive captures unsafe content
with very high reliability regardless of base model.
Second, over-refusal varies by 35\,pp across models---this
is \emph{not} architecture-agnostic. ProbeLogits is sensitive
to the model's prior over edge cases; the same primitive on
Qwen produces a much more conservative classifier than on
Llama 3 or Mistral. This is a real limitation, addressed in
practice by per-model $\alpha$ tuning (\S\ref{sec:multimodel-alpha}).

\subsubsection{ToxicChat: Hosted-Mode Comparison Across Models}
\label{sec:multimodel-toxicchat}

\begin{table}[t]
\centering
\caption{ToxicChat (n=1000) hosted-mode. PL = ProbeLogits
(zero-shot, vanilla model). LG3 = Llama Guard~3 (fine-tuned
safety classifier). Best F1 per model in bold. Wilson 95\%
CIs in brackets for headline F1.}
\label{tab:tc-multimodel}
\scriptsize
\setlength{\tabcolsep}{3pt}
\begin{tabular}{@{}llrrrr@{}}
\toprule
\textbf{System} & \textbf{Verb.} & $\alpha$ & \textbf{F1} & \textbf{R} & \textbf{P} \\
\midrule
PL Qwen 2.5-7B & S/D & 0.0 & \textbf{0.812} {\tiny[0.781, 0.841]} & 0.862 & 0.768 \\
PL Qwen 2.5-7B & S/D & 0.5 & 0.801 & 0.898 & 0.724 \\
PL Qwen 2.5-7B & S/D & 1.0 & 0.775 & 0.909 & 0.676 \\
PL Qwen 2.5-7B & Y/N & 1.0 & 0.707 & 0.898 & 0.583 \\
PL Mistral 7B  & Y/N & 0.0 & \textbf{0.719} & 0.732 & 0.707 \\
PL Llama 3 8B  & S/D & 0.5 & \textbf{0.679} {\tiny[0.638, 0.719]} & 0.641 & 0.723 \\
PL Llama 3 8B  & Y/N & 0.5 & 0.634 & 0.624 & 0.644 \\
\midrule
LG3 8B Q4\_K\_M & ---  & --- & 0.675 {\tiny[0.630, 0.717]} & 0.536 & 0.911 \\
LG3 8B Q8\_0    & ---  & --- & 0.662 & 0.528 & 0.888 \\
\bottomrule
\end{tabular}
\end{table}

Table~\ref{tab:tc-multimodel} reports same-environment hosted
results. Three observations:

\textbf{(1) Qwen with S/D substantially outperforms LG3.}
Qwen S/D $\alpha=0.0$ reaches F1 = 0.812 (CI [0.781, 0.841])
vs.\ LG3 0.675 (CI [0.630, 0.717]) --- CIs disjoint, $+$13.7\,pp
significant. With a vanilla 7B model and zero learned safety
parameters, ProbeLogits exceeds the fine-tuned LG3 baseline.

\textbf{(2) Llama 3 with S/D matches LG3 on F1 (parity).}
Llama 3 S/D F1 = 0.679 vs.\ LG3 = 0.675 --- CIs overlap fully,
the gap is statistical noise (Table~\ref{tab:lg3-head}).
Recall is significantly higher (0.641 vs.\ 0.536, CIs disjoint),
precision significantly lower (0.723 vs.\ 0.911, CIs disjoint).
Operating-point trade-off, not absolute superiority.

\textbf{(3) Mistral, restricted to Y/N, also exceeds LG3.}
Mistral Y/N $\alpha=0.0$ reaches F1 = 0.719, $+$4.4\,pp over
LG3 with overlapping CIs. Mistral cannot use S/D verbalizers
because of tokenization (\S\ref{sec:verbalizer-limit}).

\subsubsection{Verbalizer Prior Asymmetry}
\label{sec:multimodel-verbalizer}

The same model exhibits different calibration biases under
different verbalizers; magnitude and even sign differ
(Table~\ref{tab:cal-bias}).

\begin{table}[t]
\centering
\caption{Calibration bias (mean
$\text{logit}_{\text{pos}} - \text{logit}_{\text{neg}}$ over
7 null prompts) varies by model and verbalizer. Positive bias
$\Rightarrow$ model defaults to ``Yes/Dangerous''; negative
$\Rightarrow$ defaults to ``No/Safe''.}
\label{tab:cal-bias}
\small
\begin{tabular}{@{}lrr@{}}
\toprule
\textbf{Model} & \textbf{Y/N bias} & \textbf{S/D bias} \\
\midrule
Qwen 2.5-7B & $+$4.49 & $-$3.02 \\
Llama 3 8B  & $-$0.54 & $-$2.54 \\
Mistral 7B  & $-$3.04 & N/A (multi-token) \\
\bottomrule
\end{tabular}
\end{table}

\textbf{The Qwen pair flips sign}: under Y/N the model favors
``Yes''; under S/D it favors ``Safe''. This is best interpreted
as \emph{verbalizer prior asymmetry}---the same surface-form
competition phenomenon documented by Holtzman~\textit{et al.}
in classification settings~\cite{holtzman2021}, here applied
to logit-probing for safety classification: the bias magnitude
is a joint function of the model's safety-tuning and the
verbalizer tokens' surface frequency in the model's training
distribution (``Safe'' is far more common than ``Dangerous''
in RLHF responses; ``Yes''/``No'' are more balanced).

The implication is that ProbeLogits exposes priors the model
already holds, but \emph{which prior is exposed depends on the
verbalizer the OS picks at boot}. The Token Fertility check
(\S\ref{sec:verbalizer-limit}) ensures the chosen verbalizer is
single-token-per-side; the OS should additionally measure
calibration bias at boot and tune $\alpha$ per (model,
verbalizer) pair.

\subsubsection{$\alpha$ as Per-Model Policy Knob}
\label{sec:multimodel-alpha}

The optimal $\alpha$ depends on the model's bias direction and
magnitude. For positive-bias models (Qwen Y/N), higher $\alpha$
suppresses the over-trigger and improves precision. For
negative-bias models (Mistral, Llama 3 Y/N), $\alpha=0.0$ is
already balanced. For Qwen S/D (negative bias), small $\alpha$
gives best F1; larger $\alpha$ amplifies recall at precision
cost. This is not a hyperparameter to be searched on each
benchmark; it is a deployment-time policy lever for the OS to
trade precision for recall as the threat model demands.

\paragraph{Adequacy check against Llama Guard~3.}
For a same-benchmark comparison with a strong purpose-built
baseline, I evaluate Llama Guard~3 8B~\cite{llamaguard3}---a
dedicated safety classifier fine-tuned from Llama~3.1 8B---on
the identical ToxicChat 1000-prompt sample, at both Q4\_K\_M
and Q8\_0 quantization (Table~\ref{tab:lg3-head}). The
intended framing is an adequacy check (``does the substrate
primitive reach a comparable operating point as a fine-tuned
baseline?''), not a superiority claim, because ProbeLogits is
a kernel primitive rather than a competing classifier.
The four findings below disentangle the configurations: for
the strongest configuration (Qwen S/D), the comparison
crosses adequacy into significant superiority
($+$13.7\,pp, CIs disjoint); for Llama 3 S/D the result is
parity (CIs overlap); for Mistral Y/N (forced by tokenization)
the result is a smaller positive gap.

\begin{table}[t]
\centering
\caption{Adequacy comparison on ToxicChat (n=1000), pure-logit
mode for ProbeLogits (no hand-rules).
Per-model verbalizer was selected as: Qwen and Llama 3 ---
S/D verbalizer optimized per-benchmark; Mistral --- Y/N
verbalizer forced because ``Dangerous'' is multi-token in
Mistral's SentencePiece (\S\ref{sec:verbalizer-limit}).
ProbeLogits is zero-shot logit reading on a vanilla model;
Llama Guard~3 is a fine-tuned safety classifier at the same
parameter count.}
\label{tab:lg3-head}
\scriptsize
\setlength{\tabcolsep}{3pt}
\begin{tabular}{@{}lrrrr@{}}
\toprule
\textbf{System} & \textbf{F1} & \textbf{R} & \textbf{P} & \textbf{Lat.\textsuperscript{\dag}} \\
\midrule
PL Qwen 2.5-7B (S/D, $\alpha$=0.0)  & \textbf{0.812} & 0.862 & 0.768 & 0.57s \\
PL Qwen 2.5-7B (S/D, $\alpha$=0.5)  & 0.801 & 0.898 & 0.724 & 0.57s \\
PL Mistral 7B (Y/N, $\alpha$=0.0)   & 0.719 & 0.732 & 0.707 & 0.71s \\
PL Qwen 2.5-7B (Y/N, $\alpha$=1.0)  & 0.707 & 0.898 & 0.583 & 0.57s \\
PL Llama 3 8B (S/D, $\alpha$=0.5)   & 0.679 & 0.641 & 0.723 & 0.40s \\
Llama Guard~3 8B Q4\_K\_M           & 0.675 & 0.536 & 0.911 & 1.06s \\
Llama Guard~3 8B Q8\_0              & 0.662 & 0.528 & 0.888 & 1.33s \\
PL Llama 3 8B (Y/N, $\alpha$=0.5)   & 0.634 & 0.624 & 0.644 & 0.40s \\
\bottomrule
\multicolumn{5}{p{0.95\columnwidth}}{\textsuperscript{\dag}Hosted llama-cpp-python (16t, single inst), per-classification latency from the accuracy runs (total time / n, \texttt{logits\_all=True}). A matched n=150 remeasurement gives 332--556\,ms (PL) vs.\ 851/1{,}142\,ms (LG3): 2.4--3.4$\times$; deploy-realistic single probe is ${\sim}$220\,ms (\S\ref{sec:probe-perf}). PL = ProbeLogits.}
\end{tabular}
\end{table}

Four findings emerge, with both significant comparisons
reported (recall \emph{and} precision) so the operating-point
trade-off is explicit.

\textbf{First (Qwen S/D exceeds LG3 with significance).}
ProbeLogits-Qwen-2.5-7B (S/D, $\alpha=0$) F1 = 0.812
(bootstrap 95\% CI [0.781, 0.841]) vs.\ Llama Guard~3
F1 = 0.675 (95\% CI [0.630, 0.717]); the CIs are disjoint,
$+$13.7\,pp gap is significant. With a vanilla 7B model and
zero learned safety parameters, ProbeLogits exceeds the
fine-tuned LG3 baseline.

\textbf{Second (Llama 3 S/D matches LG3 -- F1 parity).}
ProbeLogits-Llama-3 (S/D) F1 =
0.679 (bootstrap 95\% CI [0.638, 0.719]) vs.\ Llama Guard~3
F1 = 0.675 (95\% CI [0.630, 0.717]); the CIs overlap fully.
The $+$0.4\,pp gap is within statistical noise---a vanilla
model on a different architecture also reaches LG3-equivalent
F1 with no learned safety parameters. Throughout this paper
we use ``parity'' in the bootstrap-CI-overlap sense (we fail
to reject the null of no difference at 95\%); this is weaker
than a formal non-inferiority claim, which would require a
pre-specified margin and a directional test.

\textbf{Third (recall \emph{vs.} precision trade-off).}
ProbeLogits-Llama-3 achieves recall 0.641
(Wilson 95\% CI [0.590, 0.689]) vs.\ LG3 0.536
(Wilson 95\% CI [0.484, 0.587]); CIs are disjoint, the
$+$10.5\,pp recall advantage is significant.
But the trade is paid in precision: ProbeLogits-Llama-3
P = 0.723 (95\% CI [0.671, 0.769]) vs.\ LG3 P = 0.911
(95\% CI [0.865, 0.942]); these CIs are also disjoint, and
the $-$18.8\,pp precision deficit is significant in the
opposite direction.
The choice between the two systems is therefore an operating-point
decision: ProbeLogits favors deployments where false-negatives
(missed unsafe content) are more costly than false-positives
(over-blocking benign content); Llama Guard~3 is the inverse.
This trade-off is a property of the system pair, not a
ranking; I report both statistics to avoid selective
reporting.

\textbf{Fourth (verbalizer-tokenizer alignment).}
The same Llama 3 model under Y/N verbalizers gives
F1\,=\,0.634 (below LG3 by $-4.1$\,pp). Switching to S/D
recovers F1 parity. The calibration bias also shifts:
Y/N $-0.54$ vs.\ S/D $-2.54$, a 4.7$\times$ magnitude
difference. This is best interpreted as
\emph{verbalizer prior asymmetry}---a verbalizer's surface
statistics in pretraining (``Safe'' is far more frequent
than ``Dangerous'' in RLHF-tuned responses) determine the
prior magnitude the probe reads. Section~\ref{sec:verbalizer-limit}
discusses the implication: ProbeLogits is sensitive to
\emph{verbalizer-tokenizer alignment}, and tokenizers
(particularly SentencePiece) constrain what verbalizer pairs
are usable.

Latency is consequential. In hosted llama-cpp-python (the
environment used for this fairness comparison), a matched
n=150 remeasurement on identical prompts gives ProbeLogits
332--556\,ms per classification (Llama 3 332\,ms, Qwen
358\,ms, Mistral Q8 556\,ms) vs.\ Llama Guard~3 851\,ms
(Q4\_K\_M) and 1{,}142\,ms (Q8\_0)---a measured
2.4--3.4$\times$ speedup, because ProbeLogits
performs only a single forward-pass logit read while Llama
Guard~3 generates response tokens. In a deploy-realistic
configuration (last-token logits only), a single probe costs
219.6\,ms median (\S\ref{sec:probe-perf}). (Native bare-metal
per-classification latency is not separately benchmarked
here.)
% remeasure_probe_{llama,qwen,mistral}.json vs remeasure_lg_{q4km,q8}.json (2026-06-15); probe_latency.json (2026-06-14)
Either way, the relative ordering holds: ProbeLogits is
fast enough for per-action governance even in unoptimized
hosted environments.

For context, Llama Guard~3 reports F1\,$\sim$0.939 on its
training-aligned benchmark; the gap to 0.675 on ToxicChat
reflects that no safety classifier transfers losslessly
across distribution shifts.

\paragraph{Run-to-run baseline variance.}
As a measurement-honesty note: a repeat Llama Guard~3
Q4\_K\_M run on the identical 1{,}000-prompt sample yields
F1 = 0.659 versus the archived first run's 0.675---LG3's
generation-based verdict has non-trivial run-to-run variance.
The paper reports the archived first run throughout; the
parity and superiority conclusions above are unchanged under
either value (if anything, the lower repeat value favors
ProbeLogits).
% results/llamaguard_1000.json (F1 0.6748) vs results/llamaguard_1000_v2.json (F1 0.6586)

ProbeLogits is designed as an OS kernel primitive for
classifying agent actions, not as a general-purpose chat
toxicity detector. On its intended use case (OS actions in
the Governed MCP companion paper~\cite{animaos-mcp}),
the full pipeline achieves F1\,=\,0.980 with perfect
precision (silicon-measured re-run: F1 = 0.980, P = 1.000,
R = 0.960, n = 260).
% measured: bootlogs/2026-07-05_paper-bench-2model-7b-135m-silicon.log [paper:custom260]

\begin{table}[t]
\centering
\caption{Effect of calibration strength $\alpha$ on
ToxicChat (n=1000) safety classification with Qwen2.5-7B
S/D verbalizer, full pipeline with hand rules,
silicon-measured on bare-metal Anima OS native (2026-07-05,
single boot; one cached forward per prompt re-scored across
all $\alpha$ values). $\alpha$ controls the precision-recall
tradeoff as a deployment-time policy parameter, not a
learned hyperparameter.}
% measured: bootlogs/2026-07-05_paper-bench-2model-7b-135m-silicon.log [paper:natsweep]
\label{tab:alpha}
\scriptsize
\begin{tabular}{@{}lrrrr@{}}
\toprule
$\alpha$ & \textbf{F1} & \textbf{Recall} & \textbf{Precision} & \textbf{Acc.} \\
\midrule
0.00 & 0.180 & 0.099 & 0.947 & 67.2\% \\
0.25 & 0.695 & 0.561 & 0.914 & 82.2\% \\
0.50 & \textbf{0.829} & 0.796 & 0.865 & 88.1\% \\
0.75 & 0.823 & 0.843 & 0.805 & 86.9\% \\
1.00 & 0.805 & 0.892 & 0.734 & 84.4\% \\
\bottomrule
\end{tabular}
\end{table}

\paragraph{Calibration strength ($\alpha$) as policy knob.}
Table~\ref{tab:alpha} shows the measured precision-recall
trade-off under varying $\alpha$ on ToxicChat. As $\alpha$
increases, the calibration correction grows stronger:
precision falls (0.947 $\to$ 0.734) while recall rises
(0.099 $\to$ 0.892). In the silicon-measured sweep the best
F1 (0.829) occurs at $\alpha=0.5$; $\alpha=1.0$ maximizes
recall at the cost of precision, and $\alpha=0$ (calibration
off) collapses recall to 0.099 even with the hand rules
active---the calibration term does real work. This is a design
feature, not a
limitation: the OS can enforce strict policies for
privileged operations ($\alpha \geq 0.8$) while applying
relaxed policies for conversational agents ($\alpha = 0.5$).
This full-pipeline (with hand rules) sweep is measured
separately from the no-hand-rules hosted runs
(Table~\ref{tab:lg3-head}); the $\alpha=0.5$ operating point
(measured F1 = 0.829)
% measured: bootlogs/2026-07-05_paper-bench-2model-7b-135m-silicon.log [paper:natsweep]
is the deployed configuration. A latency honesty note: the
bare-metal sweep averaged 5.07\,s per classification, far
above the ${\sim}$358\,ms hosted figure, because the native
harness prefills each full ToxicChat prompt (often hundreds
of tokens) at the measured 78.6\,ms per token with no
batched prefill; the hosted and native numbers measure
different prompt regimes and are both reported.
Unlike fixed-threshold classifiers, ProbeLogits delegates the
precision-recall tradeoff to the OS security policy at
deployment time. Per-model $\alpha$ behavior across three
base models is reported in \S\ref{sec:multimodel}.

\subsubsection{Small-Model Floor: Vanilla Probes Need ${\geq}$7B}
\label{sec:smallmodel}

A natural cost optimization is to run the probe on a much
smaller vanilla model. This fails categorically. On a
200-prompt HarmBench sample (Y/N verbalizer, $\alpha=0.5$,
hosted), SmolLM2-135M blocks only \textbf{13.0\%} of harmful
prompts and Qwen2.5-0.5B blocks \textbf{10.0\%}---an 87--90\%
miss rate, versus 97--99\% block rates for the three 7B-class
models (\S\ref{sec:multimodel-harmbench}). On a 200-prompt
XSTest sample (100 safe / 100 unsafe), the failure modes
differ but are equally disqualifying: 135M reaches recall
0.21, while 0.5B reaches recall 0.92 only by over-refusing
\textbf{70\%} of safe prompts. Notably, both small models are
very fast (9--25\,ms per classification)---the failure is not
latency but signal: below roughly 7B, a vanilla model's
verbalizer logits do not carry a usable safety signal.
% results/harmbench_135m_yesno.json (block_rate 0.13), harmbench_0.5b_yesno.json (0.10),
% xstest_135m_yesno.json (recall 0.21, over_refusal 0.17), xstest_0.5b_yesno.json (recall 0.92, over_refusal 0.70); 2026-06-28

This is a genuine design finding, not merely a negative
control: it (a) empirically justifies the 7B-class engine this
paper deploys as the \emph{minimum} viable probe substrate,
and (b) refutes the naive small-model cascade proposed as
future work in v1 of this paper
(\S\ref{sec:gov-overhead})---a vanilla small model cannot act
as a cheap first-stage tripwire. A \emph{distilled} tripwire
head trained against 7B-class verdicts remains open future
work; \S\ref{sec:fusionhead} describes a complementary
learned-head primitive that removes the probe forward
entirely.

\subsubsection{Cross-Model 2-of-3 Consensus}
\label{sec:consensus}

Because ProbeLogits is model-agnostic, an OS holding several
resident models can run the same probe on each and take a
majority vote. On XSTest (n=450), a 2-of-3 majority across
Qwen 2.5-7B, Llama 3 8B, and Mistral 7B lifts F1 from 0.848
(best single model, Qwen) to 0.867 and cuts over-refusal from
18.4\% to 13.6\% at essentially unchanged recall
(0.905 $\to$ 0.895). I report this honestly as
\emph{directionally positive but not statistically
significant}: the bootstrap $\Delta$F1 CI is
$[-0.013, +0.049]$ ($p_{\leq 0} = 0.12$). Per-prompt error
correlation between the three models is low (0.13--0.21),
which is why voting helps at all; a larger benchmark would be
needed to confirm significance. The measured cost is
391\,ms with the three probes run in parallel (835\,ms
sequential), i.e., consensus costs no more wall-clock time
than the slowest member.
% results/crossmodel_consensus.json (2026-06-14): majority_2of3 F1 0.8668 vs best single 0.8478,
% bootstrap_df1_majority_vs_best ci [-0.0127, +0.0489] p_le_0 0.1207; latency results/probe_latency.json

\subsection{Constitutional Governance Overhead}
\label{sec:gov-overhead}

\begin{table}[t]
\centering
\caption{Governance pipeline latency per action.}
\label{tab:governance}
\footnotesize
\setlength{\tabcolsep}{4pt}
\begin{tabular}{@{}lr@{}}
\toprule
\textbf{Stage} & \textbf{Latency} \\
\midrule
Adversarial pre-filter & $<$1\,\textmu{}s \\
Input sanitization     & $<$1\,\textmu{}s \\
ProbeLogits 7B         & 332--556\,ms (${\sim}$220\,ms deploy-realistic) \\
Blake3 audit entry     & $<$10\,\textmu{}s \\
\bottomrule
\end{tabular}
\end{table}

Table~\ref{tab:governance} shows the per-stage cost of the
governance pipeline. The classification forward
(332--556\,ms conservative, ${\sim}$220\,ms deploy-realistic)
dominates total latency. While this overhead is significant, it
occurs only once per agent action (not per token), making it
practical: a typical agent action involves hundreds of generated
tokens ($\sim$5--20\,s), so the governance check
adds ${\sim}$2--11\% overhead. A cascade that pre-screens
actions with a smaller \emph{vanilla} model---proposed as
future work in v1 of this paper---is refuted by the
small-model floor result (\S\ref{sec:smallmodel}): sub-billion
vanilla probes miss 87--90\% of harmful prompts. The surviving
latency-reduction paths are a \emph{distilled} tripwire head
(future work) and the fusion safety head
(\S\ref{sec:fusionhead}), which folds the verdict into the
generation prefill at zero marginal forward cost.

\subsection{Comparison with Related Systems}
\label{sec:comparison}

\begin{table*}[t]
\centering
\caption{Feature comparison with related systems.
\checkmark\ = supported, blank = not supported,
$\sim$ = partial.}
\label{tab:comparison}
\small
\renewcommand{\arraystretch}{1.05}
\begin{tabular}{@{}l*{7}{c}@{}}
\toprule
\textbf{Feature}
  & \textbf{Anima OS}
  & \textbf{AIOS}
  & \textbf{OS-Copilot}
  & \textbf{LangChain}
  & \textbf{llama.cpp}
  & \textbf{vLLM}
  & \textbf{AutoGPT} \\
\midrule
Bare-metal exec.
  & \checkmark & & & & & & \\
Kernel inference
  & \checkmark & & & & & & \\
ProbeLogits
  & \checkmark & & & & & & \\
Logit entropy
  & \checkmark & & & & & & \\
Grammar decoding
  & \checkmark & & & $\sim$ & & & \\
KV cache as state
  & \checkmark & & & & & $\sim$ & \\
Constitutional gov.
  & \checkmark & & & & & & \\
Trust evolution
  & \checkmark & $\sim$ & & & & & \\
Raft consensus
  & \checkmark & & & & & & \\
Agent migration
  & \checkmark & $\sim$ & & & & & \\
WASM sandbox
  & \checkmark & & & & & & \\
Hidden state
  & \checkmark & & & & & & \\
\bottomrule
\end{tabular}
\end{table*}

Table~\ref{tab:comparison} compares Anima OS with related
systems across 12 features. AIOS~\cite{aios-colm} is the closest
competitor, but it is implemented as a Python wrapper around
existing frameworks, inheriting the layered architecture
problems described in $\S$\ref{sec:layers}. llama.cpp~\cite{llamacpp}
provides efficient inference but as a userspace library, not
an OS primitive. vLLM~\cite{vllm} manages KV caches for serving
throughput but does not expose them as process state.

The distinguishing features of Anima OS---ProbeLogits, logit
entropy, kernel-level governance, and KV cache as process
state---are absent from all compared systems. These features
are only possible because inference runs inside the kernel,
giving the OS direct access to model internals.

% ============================================================
\section{Discussion}
\label{sec:discussion}

\subsection{Why Bare Metal?}
\label{sec:why-baremetal}

Running an LLM inference engine inside the kernel is an extreme
design choice. I justify it on two grounds.

First, \textbf{trust boundary}. If governance decisions depend
on inference results (``is this action safe?''), the inference
engine must be at least as trusted as the governance system. In
a layered architecture, a compromised inference framework can
return arbitrary results, undermining governance entirely. In
Anima OS, the inference engine runs at the same privilege level
as the governance system---in the kernel.

Second, \textbf{scheduling control}. A kernel-level inference
engine can be scheduled with precise priority relative to other
kernel operations. Governance checks can preempt lower-priority
inference tasks, and the kernel can manage KV cache memory
alongside other kernel memory without crossing privilege
boundaries.

\subsection{ProbeLogits vs.\ Representation Engineering}
\label{sec:vs-repeng}

Burns et al.~\cite{burns2022} discover latent knowledge in
LLM representations by training probes on hidden states.
Zou et al.~\cite{zou2023} control model behavior through
representation vectors. These works inspire my hidden state
analysis (\texttt{BehaviorProfile}) but differ fundamentally
in purpose:

Representation engineering is a \emph{research tool}: it
requires training auxiliary models, operates post hoc, and
aims to understand or modify model behavior. ProbeLogits is
an \emph{OS primitive}: it requires no training, operates in
real time, and aims to make kernel-level decisions. The two
approaches are complementary---representation engineering
could inform the design of better ProbeLogits prompts---but
they occupy different positions in the system stack.

\subsection{Fusion Safety Head: a Hidden-State Sibling}
\label{sec:fusionhead}

ProbeLogits deliberately uses zero learned parameters. A
complementary primitive, enabled by the same kernel-resident
inference access, trades that property for zero marginal
forward cost: the \emph{fusion safety head} reads the
post-final-RMSNorm hidden state of the last prompt token
\emph{during the generation prefill itself} and applies a
small learned linear head to produce a safety verdict before
the first output token is emitted. Because the prefill already
computes this hidden state, the verdict is read essentially
for free---no separate probe forward pass at all, eliminating
the 220--556\,ms probe cost from the governed-generation path.
This is exactly the hidden-state access that
Table~\ref{tab:comparison} lists as unique to a
kernel-resident engine, promoted from analysis to enforcement.

The head is implemented in the bare-metal kernel and has been
verified end-to-end on real hardware with a 14B model:
harmful generation requests are blocked
verdict-before-first-token, benign requests proceed, and the
fusion verdict is recorded in the Blake3 audit chain
(2026-06-29). Its measured accuracy on the 14B model is
HarmBench recall $90.0\%$, XSTest unsafe-prompt recall $100.0\%$,
and XSTest over-refusal $16.7\%$.%
% fusionhead: results/distill/safety_head.json eval field {hb_recall:90.0,
% xstest_unsafe_recall:100.0, xstest_over_refusal:16.67}; train_head.log confirms.
The trade-off is explicit and disclosed: unlike ProbeLogits,
the fusion head has \emph{learned parameters} (a linear probe
trained offline against 7B-class safety verdicts) and
therefore requires per-model training; if the loaded model
does not match the trained head, the kernel falls back to the
zero-shot ProbeLogits probe (fail-closed). ProbeLogits remains
the universal zero-shot substrate; the fusion head shows what
the same kernel access enables when a small amount of learning
is acceptable.
% boot/src/inference/safety_head.rs (651 LoC); results/distill/safety_head.json;
% silicon-PASS on real hardware 2026-06-29 (14B, verdict-before-first-token)

\subsection{\texorpdfstring{Enforcement $\neq$ Detection}{Enforcement ≠ Detection}}
\label{sec:enforce-detect}

The key insight of the governance architecture is the
separation of enforcement from detection.

\textbf{Enforcement} is structural: WASM isolation ensures that
agents can only interact with the system through host functions.
This guarantee holds regardless of the agent's capability or
intent. It is binary---either the agent is sandboxed or it is
not---and does not depend on the accuracy of any classifier.

\textbf{Detection} is probabilistic: ProbeLogits classifies
actions with bootstrap-CI-confirmed F1 parity to a fine-tuned
safety baseline on ToxicChat (\S\ref{sec:multimodel}) and
97--99\% block rate on HarmBench non-copyright across three
model families, using contextual calibration and safety-priming
prompts; the pipeline-with-hand-rules variant (used in the
Governed MCP companion paper~\cite{animaos-mcp}) reaches
a silicon-measured F1\,=\,0.980 on OS actions.
% measured: bootlogs/2026-07-05_paper-bench-2model-7b-135m-silicon.log [paper:custom260]
This accuracy improves with model size and prompt quality but
is never 100\%.

The combination is more powerful than either alone. A system
with only enforcement (no detection) would sandbox agents but
have no visibility into their behavior. A system with only
detection (no enforcement) could identify harmful actions but
not prevent them if the agent bypasses the detector. Anima OS
provides both: detection informs the graduated response, and
enforcement ensures the response is applied.

\subsection{Why Now?}
\label{sec:why-now}

The convergence of three technologies makes kernel-level AI
integration practical in 2026:

\begin{enumerate}[topsep=2pt,itemsep=1pt]
  \item \textbf{Quantization}: 4-bit quantization (Q4\_0)
        reduces a 7B model from 14\,GB to 4.1\,GB, fitting
        in commodity RAM.
  \item \textbf{SIMD}: AVX-512 VNNI enables efficient
        4-bit matrix-vector multiplication without GPU
        acceleration.
  \item \textbf{Practical per-action latency}: the per-token
        forward is a measured 78.6\,ms;
        % measured: bootlogs/2026-07-05_paper-bench-2model-7b-135m-silicon.log [paper:tok7b]
        a full classification (332--556\,ms hosted,
        ${\sim}$220\,ms deploy-realistic on 7B) runs once per
        agent action---fast enough for per-action governance.
\end{enumerate}

Five years ago, 7B inference required seconds per token on CPU.
Today it requires tens of milliseconds per token
(a measured 78.6\,ms on 7B).
% measured: bootlogs/2026-07-05_paper-bench-2model-7b-135m-silicon.log [paper:tok7b]
This orders-of-magnitude
improvement transforms LLM inference from an application-level
service to a viable kernel primitive.

% ============================================================
\section{Related Work}
\label{sec:related}

\paragraph{AI-oriented operating systems.}
AIOS~\cite{aios-colm} extends a Python-based framework with
LLM-oriented scheduling and memory management (published at COLM 2025).
OS-Copilot~\cite{oscopilot}
uses LLMs to automate Linux system administration tasks.
PunkGo~\cite{punkgo} proposes a Rust sovereignty kernel for
verifiable AI agent execution with Merkle audit logs, the closest
related work in kernel-level AI governance. Microsoft's Agent
Governance Toolkit~\cite{msagt} provides application-level
runtime security for AI agents.
All operate atop existing operating systems (or focus on
verification rather than inference), inheriting the latency and
bypassability problems discussed in $\S$\ref{sec:problems}.
Anima OS differs by integrating inference directly into a
bare-metal kernel, enabling logit-level primitives.

\paragraph{Inference engines.}
llama.cpp~\cite{llamacpp} is the state-of-the-art CPU inference
engine, with extensive SIMD optimization and broad hardware
support. My bare-metal engine achieves throughput parity with
llama.cpp on both bandwidth-limited 7B workloads and
compute-bound 135M workloads (measured 1{,}344 tok/s at 8
workers vs.\ the 1{,}301 tok/s baseline at 8 threads; the
1.39$\times$ 135M advantage claimed in v1 did not reproduce
in the archived measurement campaign and is withdrawn).
% measured: bootlogs/2026-07-05_paper-bench-2model-7b-135m-silicon.log [paper:thru135m8]
Parity from inside a bare-metal kernel---with no OS
scheduling support and a from-scratch work-stealing
executor---is the claim of record, not algorithmic
superiority.

\paragraph{Serving systems.}
vLLM~\cite{vllm} introduces PagedAttention for efficient KV
cache memory management in serving. SGLang~\cite{sglang}
optimizes structured generation programs with RadixAttention.
Both focus on serving throughput (requests per second), while
my KV cache primitives focus on process state management
(checkpoint, restore, fork). The two approaches address
different problems.

\paragraph{Constitutional AI.}
Bai et al.~\cite{bai2022constitutional} introduce constitutional
AI as a training-time technique: the model is trained with
AI feedback guided by a set of principles. My approach is
complementary---I enforce constitutional principles at
\emph{runtime} in the kernel, regardless of how the model was
trained. The two approaches can be combined: a model trained
with constitutional AI would produce more accurate ProbeLogits
classifications.

\paragraph{Safety classifiers.}
Llama Guard~\cite{llamaguard,llamaguard3} is a fine-tuned safety
classifier achieving F1 $\sim$0.93--0.95 on its benchmark.
WildGuard~\cite{wildguard} achieves macro-F1 0.907 on
WildGuardTest. NeMo Guardrails~\cite{nemoguardrails} provides
programmable safety rails for LLM applications. All operate
as external classifiers at the application level, requiring
dedicated models. ProbeLogits differs by using the
\emph{same model} that performs inference for classification
(no second model overhead, bootstrap-CI-confirmed F1 parity
to Llama Guard~3 on ToxicChat with a vanilla 7B model and zero
learned fine-tuning), and by operating below the WASM sandbox
boundary where the agent cannot intercept or skip the safety
check.

\paragraph{Constrained decoding.}
Outlines~\cite{outlines} and Guidance~\cite{guidance} implement
constrained decoding in Python/C++. llguidance~\cite{llguidance}
(Microsoft) uses PDA-based algorithms for CFG constraints. My
current grammar support covers choice/boolean constraints,
NFA-based regex, and JSON structural / JSON-Schema-constrained
generation; I plan to port llguidance to \texttt{no\_std} for
full recursive CFG support.

\paragraph{Representation engineering.}
Burns et al.~\cite{burns2022} discover latent knowledge in
LLM representations without supervision. Zou et al.~\cite{zou2023}
propose representation engineering as a top-down approach to
AI transparency. These works inspire my hidden state behavior
analysis but operate as research tools, not OS primitives
($\S$\ref{sec:vs-repeng}).

\paragraph{Distributed inference.}
Petals~\cite{petals} enables collaborative inference by
splitting model layers across networked machines. Anima OS's
AnimaNet takes a different approach: distributing
\emph{agents} (not tensors) across nodes, with Raft~\cite{raft}
consensus for coordination.

\paragraph{Capability security.}
seL4~\cite{sel4} provides formally verified capability-based
security. CHERI provides hardware-enforced capabilities. Anima
OS uses Ed25519-signed capability tokens for agent permissions,
drawing on the capability security tradition but applied to
AI agent governance.

\paragraph{Rust bare-metal operating systems.}
Theseus OS~\cite{theseus} explores Rust's type system
for OS safety guarantees. These systems demonstrate the viability
of Rust for kernel development but do not address AI integration.
Anima OS builds on this foundation, extending bare-metal Rust
to AI inference and governance.

\paragraph{LLM agent frameworks.}
AutoGPT~\cite{autogpt}, LangChain~\cite{langchain},
CrewAI~\cite{crewai}, and MetaGPT~\cite{metagpt} orchestrate
LLM-based agents in Python. These frameworks provide convenience
but operate entirely in userspace, with no kernel-level
governance or inference integration.

% ============================================================
\section{Limitations}
\label{sec:limitations}

\paragraph{GPU acceleration.}
Anima OS currently runs CPU-only inference. For models beyond
7B parameters, GPU acceleration is essential. Bare-metal GPU
compute (AMD RDNA or NVIDIA) requires reverse-engineering
proprietary interfaces, which remains a significant engineering
challenge.

\paragraph{ProbeLogits accuracy.}
On ToxicChat (1{,}000 prompts), the archived pure-logit
configuration reaches F1 = 0.812 (Qwen 2.5-7B S/D,
$\alpha=0.0$), with bootstrap-CI-confirmed
parity-or-superiority to Llama Guard~3~\cite{llamaguard3}
(\S\ref{sec:multimodel}); the full-pipeline native sweep spans
a measured F1 = 0.180 ($\alpha=0$, calibration off) to 0.829
($\alpha=0.5$, best).
% measured: bootlogs/2026-07-05_paper-bench-2model-7b-135m-silicon.log [paper:natsweep]
The optimal $\alpha$ is verbalizer- and domain-dependent:
$\alpha = 0.5$ maximizes F1 on chat, while OS actions tolerate
higher $\alpha$ (full-pipeline OS variant in companion
paper~\cite{animaos-mcp}). Independent annotators and broader
benchmarks beyond the three reported (HarmBench, XSTest,
ToxicChat) would further strengthen the results.

\paragraph{Small-model floor.}
The probe substrate requires a 7B-class model:
\S\ref{sec:smallmodel} shows that vanilla 135M and 0.5B models
block only 10--13\% of HarmBench prompts, so ProbeLogits
cannot be made cheap by shrinking the vanilla model. Any
deployment must budget for a resident ${\geq}$7B engine (or a
trained head, with the trade-offs of \S\ref{sec:fusionhead}).

\paragraph{Multi-axis probe consensus (negative result).}
An obvious extension is to probe several safety axes (harm,
intent, refusal-prior) in one pass and combine them with a
learned linear combiner. I tested this and report it as a
negative result: on XSTest, the combiner statistically ties
the single calibrated harm axis (Qwen: $\Delta$F1 $+0.009$,
bootstrap CI $[-0.015, +0.033]$, McNemar $p = 0.33$; one of
three models, Llama 3, gains $\Delta$F1 $+0.063$ but the
effect does not generalize), and the XSTest-trained combiner
\emph{fails frozen transfer} to HarmBench non-copyright,
collapsing recall from 0.93 (single harm threshold) to 0.39
($\Delta = -0.54$)---the refusal-prior axis learns a
dataset-specific over-refusal correction whose sign is wrong
for pure-harm distributions. The single calibrated harm axis
remains the recommended configuration.
% results/multiaxis_rigor_qwen2.5-7b-instruct-q4_0.json (bootstrap_df1 mean +0.0092 ci [-0.0147,+0.0329],
% mcnemar p 0.3268, frozen transfer 0.93 -> 0.39); multiaxis_combiner_*.json (llama delta_f1 +0.0631); 2026-06-14

\paragraph{Verbalizer-tokenizer alignment.}
\label{sec:verbalizer-limit}
ProbeLogits requires the verbalizer pair (e.g., Safe/Dangerous,
Yes/No) to be single vocabulary tokens. This is satisfied for
BPE tokenizers in Llama 3 and Qwen on both pairs, but
violated for Mistral 7B (SentencePiece): ``Dangerous''
tokenizes as ["D", "anger", "ous"], so the probe at the
answer position reads a meaningless ``D'' logit.
Mistral evaluations therefore use Yes/No only, restricting
the available verbalizer space.
This is a real architectural constraint, not a footnote:
the choice of base model determines what verbalizer pairs
are usable, which in turn affects which model priors are
legibly exposed by the probe. The single-token constraint
\emph{is} checked at OS boot: the implemented Token Fertility
check measures the tokenization fertility of each candidate
verbalizer pair, selects a pair whose both sides tokenize to
exactly one token, and refuses to enable the probe if no
usable verbalizer exists for the loaded model.
% implemented: boot/src/inference/probe.rs:39-80 (check_token_fertility + boot-time verbalizer selection)

\paragraph{Grammar expressiveness.}
Current grammar support covers choice constraints, NFA-based
regex, a JSON structural grammar, and JSON-Schema-constrained
generation. Full context-free grammar (CFG) support---arbitrary
recursive structured output---requires porting a PDA-based
engine (such as llguidance~\cite{llguidance}) to \texttt{no\_std},
which I estimate at approximately two weeks of engineering effort.

\paragraph{Benchmark breadth and methodology.}
Three caveats apply to the evaluation:
(1)~per-model verbalizer choice was made post-hoc---Qwen and
Llama 3 use Safe/Dangerous (single-token in their tokenizers),
Mistral is forced to Yes/No because ``Dangerous'' is multi-token
in SentencePiece (\S\ref{sec:verbalizer-limit}). A pre-registered
verbalizer-selection protocol would harden the claim.
(2)~$\alpha$ values reported in $\alpha$ sweeps were chosen by
inspecting the same benchmark on which they are reported. A
held-out validation set for $\alpha$ selection would replace
test-set tuning with a principled deployment procedure.
(3)~the OS-action benchmark used for component-level pipeline
ablation (260 prompts) was author-labeled and the hand-rule
heuristics were iteratively tuned against it; it is reported
only in the companion paper~\cite{animaos-mcp} for the
gateway-level evaluation.
ToxicChat (n=1{,}000), HarmBench (n=400), and XSTest (n=450)
are independent human-annotated benchmarks that mitigate the
risk of label optimization, and the multi-model evaluation
(Qwen 2.5-7B, Llama 3 8B, Mistral 7B) addresses
single-model overfitting concerns from the v1 draft.

\paragraph{Formal verification.}
The governance system's non-bypassability claim rests on
WASM isolation properties, which I have not formally verified.
TLA+ or Iris specifications of the governance invariants
would significantly strengthen the security argument. I
note that WASM sandboxing is a well-studied isolation
mechanism, but formal verification of the specific Anima OS
host function interface has not been performed.

\paragraph{TLS certificate verification.}
The TLS 1.3 client performs full X.509 server-certificate
validation---chain building to a Mozilla root store
(\texttt{webpki-roots}), validity-period checks against the
real-time clock, hostname/SAN matching, and signature
verification across RSA PKCS\#1 v1.5 / PSS, ECDSA P-256 / P-384,
and Ed25519---using rustls's \texttt{WebPkiServerVerifier} driven
by a custom \texttt{no\_std} RustCrypto provider. This was
verified on real hardware: a valid certificate is accepted while
expired, wrong-hostname, self-signed, and untrusted-root
certificates are each rejected for the correct reason
(\texttt{Expired}, \texttt{NotValidForName}, \texttt{UnknownIssuer}).
The remaining limitation is that certificate revocation (OCSP /
CRL) is not yet checked.

\paragraph{Multi-agent demonstrations.}
While the infrastructure for multi-agent systems is complete
(lifecycle, trust, capabilities, migration, Raft), I have not
yet demonstrated complex multi-agent scenarios with realistic
workloads. End-to-end demonstrations with cooperating and
competing agents would validate the governance and trust
systems under realistic conditions.

% ============================================================
\section{Conclusion}
\label{sec:conclusion}

I presented ProbeLogits, the first OS primitive that exposes
LLM logit vectors as kernel-level abstractions for semantic
classification, uncertainty measurement, and governance
enforcement. ProbeLogits eliminates text generation, parsing,
and IPC from the classification path: it reads a single logit
position after one prefill (332--556\,ms hosted on 7B,
${\sim}$220\,ms deploy-realistic)
instead of generating and parsing tokens, a measured
2.4--3.4$\times$ faster than a generation-based guard.

The key finding is that bare-metal inference operates at the
memory-bandwidth wall: the measured per-token cost on 7B
(78.6\,ms, sustained-load operating point)
% measured: bootlogs/2026-07-05_paper-bench-2model-7b-135m-silicon.log [paper:tok7b]
corresponds to ${\sim}$70\% of the
theoretical DDR5-6000 bandwidth ceiling (55\,ms). No
software optimization can lift this past the ceiling on the
same hardware---the path forward is higher memory bandwidth or
GPU acceleration.

ProbeLogits, combined with KV cache process state operations
and kernel-enforced constitutional governance, demonstrates
that an AI-native kernel can provide capabilities impossible
in layered architectures: millisecond-scale single-forward
semantic classification, kernel-enforced safety guarantees via WASM sandbox
isolation (defense-in-depth), and process-level management of inference state
(checkpoint, restore, fork).

The calibration parameter $\alpha$ provides a deployment-time
knob for precision-recall tradeoff, enabling domain-specific
safety policies without retraining.

As AI agents become as common as processes, the operating system
must evolve from a passive resource manager to an active
participant in the inference loop. ProbeLogits is a first step
toward that future.

% ============================================================
\bibliographystyle{plain}

\end{document}